\documentclass[%
reprint,
amsmath,amssymb,
aps,
]{revtex4-1}

\usepackage{graphicx}% Include figure files
\usepackage{dcolumn}% Align table columns on decimal point
\usepackage{bm}% bold math

\usepackage{placeins}

\usepackage{hyperref}% add hypertext capabilities
\newcommand{\Zeff}[0] {Z_{\mathrm{eff}}}
\newcommand{\mean}[1] {\langle#1\rangle}
\newcommand{\abs}[1]{\vert #1 \vert}
\newcommand{\ohm}{$\mathrm{\Omega}$~}
\newcommand{\powerten}[2]{$\mathrm{#1 \times 10^{#2}}$}
\newcommand{\refeq}[1]{Eq.(\ref{#1})}
\newcommand{\reffig}[1]{Fig.\ref{#1}}
\newcommand{\customsubsection}[1]{\emph{#1}.—}

\begin{document}

%\preprint{APS/123-QED}

\title{Fundamental and Environmental Contributions to the Cyclostationary Third Moment of Current Fluctuations in a Tunnel Junction}% Force line breaks with \\

\author{Pierre F\'evrier}
\author{Christian Lupien}
\author{Bertrand Reulet}%

\affiliation{
Institut Quantique and D\'epartement de Physique, Universit\'e de Sherbrooke, Sherbrooke, Qu\'ebec J1K 2R1,Canada
}

\date{\today}% It is always \today, today,
             %  but any date may be explicitly specified

\begin{abstract}
Current fluctuations generated by tunnel junctions are known to be non-Gaussian. However, this property is lost when fluctuations are measured at high frequency and limited bandwidth. We show that the quadratures of the electric field generated by a tunnel junction at frequency $f$ displays third order correlations, i.e. skewness, when the junction is electrically driven at $3f$, revealing the Poisonnian statistic of charge transfer by the barrier even at short time-scales. In addition to this intrinsic contribution from the junction, we observe extra correlations induced by the environmental noise at frequency $f$ as well as a feedback effects coming from the environmental impedance not only at frequency $f$ but also at some multiples of $f$.
\end{abstract}

\pacs{Valid PACS appear here}% PACS, the Physics and Astronomy
                             % Classification Scheme.
%\keywords{Suggested keywords}%Use showkeys class option if keyword
                              %display desired
\maketitle

%\tableofcontents

\section{\label{sec:Intro}Introduction}

Electronic noise in mesocopic conductors has been the object of many investigations \cite{blanterShotNoiseMesoscopic2000}. Indeed, quantum transport is determined by the statistics, dynamics and interactions of charge carriers. These are imprinted in the statistics of current fluctuations, the study of which thus provides insights into the conduction mechanisms. Many experiments deal with the variance of voltage or current fluctuations, measured over long timescales i.e. at low frequency, which already provides interesting information beyond the measurement of the conductance. However, in order to access dynamical properties, it is usually necessary to perform detection at shorter timescales i.e. to work at finite frequency $f$. Unfortunately, the variance of fluctuations, which is the simplest quantity to measure beyond average current, is usually frequency independent when charge is conserved \cite{hekkingFiniteFrequencyQuantum2006}. Beyond the variance, the noise susceptibility, i.e. the dynamical response of noise to an ac excitation, has proven to be a probe of the energy relaxation time in wires \cite{reuletNoiseThermalImpedance2005,pinsolleDirectMeasurementElectron2016a}. Photo-assisted noise can also be used to access the same quantity\cite{bagretsFrequencyDispersionPhotonassisted2007}. 

Beyond usual variance, higher order moments have been measured in various samples \cite{bomzeMeasurementCountingStatistics2005,gabelliFullCountingStatistics2009}. For systems with slow dynamics like quantum dots, frequency-dependent statistics reveal the tunnel rates through the barriers \cite{ubbelohdeMeasurementFinitefrequencyCurrent2012a,fujisawaElectronCountingSingleelectron2004}. In samples with fast dynamics such as metallic wires \cite{nagaevFrequencyScalesCurrent2004,pilgramFrequencydependentThirdCumulant2004,galaktionovStatisticsCurrentFluctuations2003}, the frequencies involved are in the GHz range, and microwave techniques are mandatory. The third moment of current fluctuations in such a frequency domain has been performed in tunnel junctions\cite{reuletEnvironmentalEffectsThird2003} and short diffusive wires\cite{pinsolleNonGaussianCurrentFluctuations2018}. These experiments face, beside the very low signal, two difficulties: first, the need for a very wide bandwidth, which is difficult to achieve in microwave circuits. Second, it involves environmental contributions which arise as soon as the impedance of the detecting apparatus is non-zero\cite{beenakkerTemperatureDependentThirdCumulant2003}. In practice, this impedance is usually $50\Omega$ and of the same order as that of the sample. As a consequence, there is no report of detection of third order fluctuations in mesoscopic devices beyond $1\mathrm{GHz}$, except by a mixed detection which involves both low and high frequencies \cite{gabelliElectronPhotonCorrelations2013,gabelliHighFrequencyDynamics2009}.

The constraint on the detection bandwidth is stringent: a signal in the range $[f_1,f_2]$ has no third moment if $f_2<2f_1$, so for example, experiments working with a 4-8 GHz bandwidth, common in the detection of cryogenic microwave signals, are useless for the detection of a third moment. This severe condition originates from stationary: the third moment in frequency domain $\langle i(f_1)i(f_2)i(f_3)\rangle$ is zero unless $f_1+f_2+f_3=0$. This condition can be relaxed into $f_1+f_2+f_3=nf_0$ with $n$ any integer if the system is excited by a periodic signal at frequency $f_0$. This condition of cyclostationarity can be obeyed with a detector of narrow bandwidth for $n =\pm1$, $f_0=f$, to give the cyclic moment $K_1(f)=\langle i^2(f)i(-f)\rangle$, or for $f_0=3f$, to give the cyclic moment $K_3(f)=\langle i^3(f)\rangle$.
Under cyclostationary conditions, a third moment of current fluctuations can in principle be measured with a narrow bandwidth detection scheme. The purpose of this article is to implement such a measurement to address the second difficulty related to measurements of third moments in the microwave domain: what are the environmental effects in this measurement? This question has been partially addressed theoretically \cite{heikkilaCyclostationaryMeasurementLowfrequency2005}.

%Plan du papier
This communication is organized as follows: in section II we describe the experimental setup and results of the measurement of the cyclostationary third moment of voltage fluctuations generated by a tunnel junction placed at low temperature; in section III we analyze the results in terms of intrinsic contributions and environmental effects. Section IV contains a conclusion, remarks and perspectives.

\section{\label{sec:Setup}Experimental setup}

\begin{figure}
 	\includegraphics[width=0.5\textwidth]{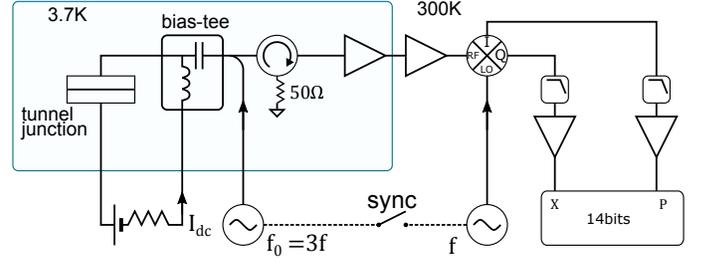}
 	\caption{\label{fig:setup} Experimental setup. The sample is excited at frequency $f_0=3f=14.55$GHz. The LO of the IQ mixer is at frequency $f=4.85$GHz.}
\end{figure}

\customsubsection{Sample and biasing}
Tunnel junctions are known to produce non-Gaussian fluctuations due to the binomial statistics of the charge transfer through the tunneling barrier. We have used a junction similar to the ones used in shot-noise thermometry\cite{spietzPrimaryElectronicThermometry2003}, with a planar geometry of area $5\times1\mathrm{\mu m}$. It has been made using usual lithography techniques and evaporation of aluminum electrodes on a silicon substrate. The insulating tunnel barrier is obtained by controlled oxidation of the first electrode under oxygen atmosphere. The dc resistance of the junction is $R_J=$130\ohm at 3.7K (116\ohm at 300K). The experiment has been performed in a helium-free cryostat with a base temperature of 3.7K so that aluminum is not superconducting. The junction is current biased through a bias-Tee (see \reffig{fig:setup}), allowing the separation of high frequency signals from the dc bias line.  The junction is ac-biased by a single tone at frequency $f_0=3f=$14.55GHz using a directional coupler.

%Signal attendu: S3 lineaire en I independant de Idc et de la temperature.

\customsubsection{Homodyne measurement}
The spectral density of voltage fluctuations at the junction is of the order of $ 10^{-10}\mathrm{V/\sqrt{Hz}}$ which is way too small for direct detection without amplification. The microwave signal generated by the junction propagates through an isolator before amplification by a low-noise 4-8GHz cryogenic amplifier with a noise temperature of 2.5K. The role of the isolator is to make the temperature and the impedance of the environment seen by the junction well defined: a 50\ohm load at 3.7K instead of the input of the amplifier with an unknown, frequency-dependent impedance and noise temperature. After extra amplification at room temperature, the signal is down-converted from $f=$4.85GHz to low frequency by an IQ mixer with high linearity. The two resulting quadratures $X$ and $P$ are amplified and low-pass filtered, with a measurement bandwidth of $\Delta f=$225MHz. The signal is then digitized with a fast acquisition card (14bits, 400MS/s). The joint probability density $\mathcal{P}(X,P)$ is calculated on the fly from the acquired data. The skewness of the voltage fluctuations on each quadrature $\mean{X^3}$ and $\mean{P^3}$ is directly linked to the cyclic third moments $K_{v,3}(f)=\mean{v_{mes}(f)^3}$ and $K_{v,1}(f)=\mean{v_{mes}(f)^2v_{mes}(-f)}$ of voltage fluctuations $v_{mes}(f)$ measured at the input of the cryogenic amplifier:
\begin{equation}
\mean{X^3}+i\mean{P^3}=\frac{3}{4}G^3\left[K_{v,3}(f)e^{i3\phi_0}+3K_{v,1}(f)e^{i\phi_0}\right]\Delta f^2
\end{equation}

\noindent where $G$ is the total voltage gain of the amplification chain and $\phi_0$ a global phase due to the delay between excitation and detection. The amplitude of the ac excitation $I_{ac}$, the gain $G$ and the noise added by the amplification chain, are calibrated by measuring the variance of the photo-assisted noise $\mean{X^2}$ and $\mean{P^2}$ vs. $I_{dc}$, the theory of which is well established \cite{lesovikNoiseAcBiased1994,schoelkopfObservationPhotonAssistedShot1998}.The power gain $G^2$ is estimated around 78dB and the effective noise temperature of the measurement is 3K, as expected for a 130\ohm load with this specific cryogenic amplifier. The signal to noise ratio at high bias is limited by the noise of the junction itself.

 \begin{figure}
	\centering
	\includegraphics[width=0.55\textwidth]{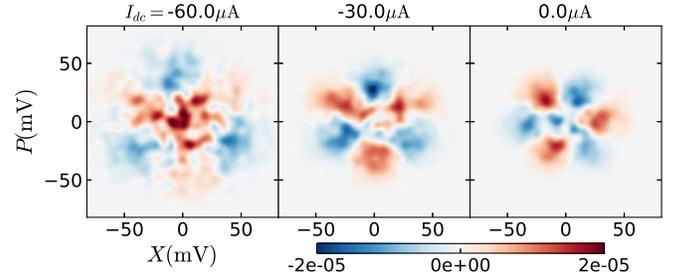}
	\caption{Probability density $\mathcal{P}(X,P)$ of the quadratures of voltage fluctuations at frequency $f=4.85$GHz generated by a tunnel junction under cyclostationary excitation at frequency $3f$ (after subtraction of the phase insensitive contributions). The existence of a third order moment is revealed by the rotational symmetry clearly visible in the data. Different plots correspond to different dc biases.}
	\label{fig:DPxy}
\end{figure}

\customsubsection{\label{sec:extrCorr}Third moment from the symmetries of the histograms}The phase coherence between the detection at frequency $f$ and the excitation at frequency $3f$ can be switched on or off. When off, a slight detuning of one of the microwave sources averages to zero the contributions that depend on $\phi_0$ and thus should lead to $\mean{X^3}=0$ and $\mean{P^3}=0$. All remaining contributions are due to the non-linearity of the amplification chain, and of the acquisition card. To remove these unwanted contributions in $\mathcal{P}(X,P)$, we measure the difference in histograms obtained with and without phase coherence. The resulting differential probability $\Delta \mathcal{P}(X,P)$, showed \reffig{fig:DPxy} has a finite order rotational symmetry. This is the direct consequence of the homodyne demodulation of the noise at a fraction of the modulation frequency. It is then natural to write $\mathcal{P}(X,P)$ in polar coordinates: $X=r\cos \theta$ and $P=r\sin \theta$ and express $\mathcal{P}(r,\theta)$ as the Fourier series:

\begin{equation}
\mathcal{P}(r,\theta)=\sum_{n\in\mathbb{Z}} \mathcal{P}_n(r)e^{-i n\theta}
\end{equation}
We define:
\begin{equation}
W_{\alpha,n}=\frac{\pi}{2}\int_0^{+\infty} \mathcal{P}_n(r)r^{\alpha+1}dr
\label{eq:defW}
\end{equation}

\noindent Moments of the probability distribution $\mathcal{P}(X,P)$  are related to the $W_{\alpha,n}$, which can be interpreted as the contribution of the rotational symmetry of order $n$ to the moment of order $\alpha$. For the third moment we find:
\begin{equation}
\begin{split}
&\mean{X^3}=\mathrm{Re}(W_{3,3}+3W_{3,1})\\
&\mean{XP^2}=\mathrm{Re}(W_{3,1}-W_{3,3})\\
&\mean{P^3}=\mathrm{Im}(W_{3,3}-3W_{3,1})\\
&\mean{PX^2}=\mathrm{Im}(W_{3,1}+W_{3,3})
\end{split}
\end{equation}

\begin{figure}
	\includegraphics[width=0.5\textwidth]{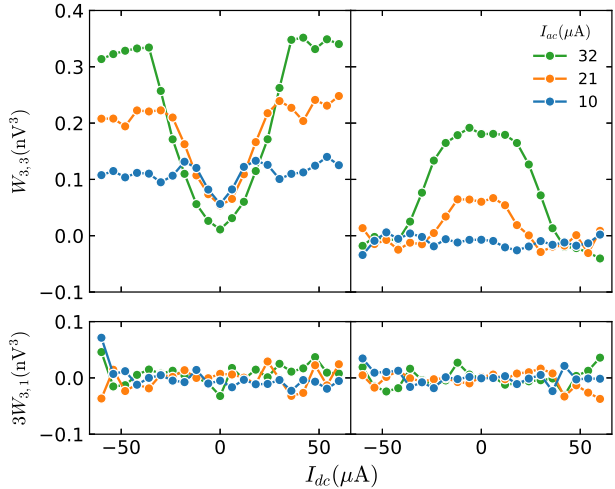}
	\caption{\label{fig:K3raw} Contributions $W_{3,3}$ and $W_{3,1}$ to the third moment of voltage fluctuations as a function of $I_{dc}$ for various ac excitations. They are calculated using \refeq{eq:defW} from the measurements of the differential probability density $\Delta \mathcal{P}(X,P)$ of \reffig{fig:DPxy}. The distributions $\mathcal{P}(X,P)$ have been rotated by a global angle in order to maximize the signal at high dc bias on the $X$ quadrature. Each data point is averaged over 45mn.}
\end{figure}

The skewness of the marginal probability distributions $\mathcal{P}(X)$ and $\mathcal{P}(P)$ is finite when $\mathcal{P}(X,P)$ shows either a one-fold or three-fold symmetry characterized respectively by $W_{3,1}$ and $W_{3,3}$. The measurement of the joint probability is necessary to separate these two contributions. In our experiment, it appears that $W_{3,1}\simeq0$ (see \reffig{fig:K3raw}). This corresponds to $K_{v,1}= 0$ as expected for a modulation at frequency $3f$. We consider in the following $\mean{X^3}=\mathrm{Re}(W_{3,3})$ and $\mean{P^3}=\mathrm{Im}(W_{3,3})$, in order to improve signal to noise ratio, and an arbitrary phase $\phi_0$ that maximizes $\mean{X^3}$ and makes $\mean{P^3}$ vanish at high bias.

From \reffig{fig:K3raw}, we see that the measured third moment at high bias ($I_{dc}>$40$\mathrm{\mu A}$) shows a plateau with an amplitude proportional to $I_{ac}$. This is qualitatively close from what is expected for a tunnel junction, i.e. a third moment proportional to the average current $\mean{I}$: $\mean{(I-\mean{I})^3}=e^2\mean{I}$ \cite{levitovCountingStatisticsTunneling2004,bomzeMeasurementCountingStatistics2005}. In the limit where the noise is adiabatically modulated, all the moments of the current distribution follow the bias modulation, and we should have $K_{v,3}\propto e^2I_{ac}$. The third moment at low bias shows however a clear deviation from this behavior. In the following we show that there are extra contributions that comes from the measurement setup and demonstrate how to separate them.

\section{\label{sec:subEnv} Effect of the environment}

\begin{figure}
	\centering
	\includegraphics[width=1\linewidth]{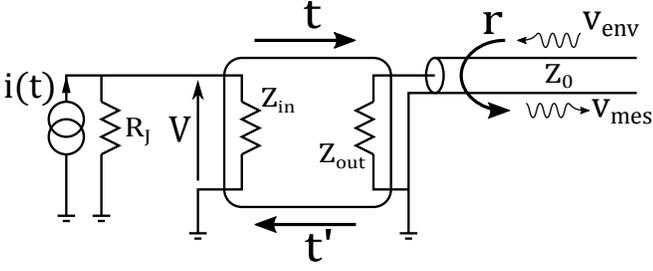}\caption{\label{fig:model}Model for the impedance mismatch between the junction and the transmission line. The box represents an effective 4 ports circuit.}
\end{figure}

Ideally, current fluctuations in a conductor should be measured using a ammeter with both a fast response and a zero input impedance. However, microwave equipment has a 50$\mathrm{\Omega}$ input impedance. Furthermore, the tunnel junction is embedded in an electromagnetic environment consisting of its own capacitance, connecting leads, wire bonds and a transmission line (TL). This can be modeled by the effective reciprocal circuit showed in \reffig{fig:model}. The input impedance $Z_{in}$ represents the impedance seen by the junction, the output impedance $Z_{out}$, the effective load on the TL. The voltage transmission coefficients $t$(resp. $t'$), from the junction to the TL (resp. the TL to the junction), include the finite propagation time in the circuit. This environment induces voltage fluctuations across the junction given by:

\begin{equation}\label{eq:v_J}
\delta V(f)= -i(f)\Zeff(f)+t'(f)v_{env}(f)
\end{equation}

\noindent with $\Zeff=R_JZ_{in}(f)/(R_J+Z_{in}(f))$. $v_{env}$ is the noise coming from the measurement setup (here dominated by the thermal noise of the 50\ohm load of the isolator) and $i(f)$ the current fluctuations generated by the junction. Voltage fluctuations across the junction lead to environmental corrections to the cyclostationary third moment, in a similar way they do in the stationary regime \cite{reuletThirdMomentCurrent2010a,beenakkerTemperatureDependentThirdCumulant2003,kindermannFeedbackElectromagneticEnvironment2004}. Using circuit theory, we find the different contributions to the cyclostationary third moment of voltage fluctuations detected by the amplifier connected to the transmission line\cite{heikkilaCyclostationaryMeasurementLowfrequency2005}:

\begin{equation}\label{eq:K3_all}
	K_{v,3}(f)=t(f)^3R_J^3\left[K_{int}(f)+K_{env}(f)+K_{fb}(f)\right]
\end{equation}

\noindent with $K_{int}$ the intrinsic contribution of the sample, $K_{env}$ the contribution due to the term $t'(f)v_{env}(f)$ in \refeq{eq:v_J} and $K_{fb}$, the contribution due to the term $i(f)\Zeff(f)$. In \cite{heikkilaCyclostationaryMeasurementLowfrequency2005}, these quantities have been calculated at zero bias and zero frequency, i.e. neglecting the signal propagation in the circuit, which does not correspond to our experiment. In the following we derive the last two terms in \refeq{eq:K3_all} at any bias $I_{dc}$, and in the case of realistic microwave setup. We have computed the contributions to the third moment due to thermal fluctuations of the environment $K_{env}$ and the feedback effect $K_{fb}$ using the so-called cascaded Langevin approach \cite{beenakkerTemperatureDependentThirdCumulant2003}, where we use the separability of timescales between the fluctuations $\delta V(t)$ and $i(t)$. 

Considering that the noise generated by the junction responds adiabatically to the applied voltage, we have:
\begin{equation}\label{eq:St}
\mean{i(t)^2}=S(V(t))+\delta V(t)\frac{dS}{dV}\Bigr|_{V(t)}
\end{equation}
\noindent where brackets $\mean{}$ designate an ensemble average and $V(t)=R_J\left[I_{dc}+I_{ac}\cos (2\pi f_0 t)\right]$ is the periodically modulated bias voltage. In the following, we choose a more general approach.
Because of cyclostationarity, Fourier components $i(f')$ separated by a frequency $\alpha f_0$ where $f_0$ is the driving frequency (here $f_0=3f$ with $f$ the detection frequency) and $\alpha$ an integer, can be correlated:
\begin{equation}
	S_\alpha(f')=\mean{i(f')i(\alpha f_0-f')}
\end{equation}

 \noindent $S_\alpha(f')$ is the dynamical response of order $\alpha > 0$ of the noise at frequency $f'$ to an excitation at $f_0$ \cite{gabelliDynamicsQuantumNoise2008}. $S_0$ is the static response, i.e. the photo-excited noise. On the top of the excitation at $f_0$, the voltage fluctuations $\delta V(f+\varepsilon)$ around frequency $f$ introduce additional correlations: 
\begin{equation}
	 \mean{i(f')i(\alpha f_0+f+\varepsilon-f')}=D_\alpha(f')\delta V(f+\varepsilon)
\end{equation}
 
\noindent where $D_\alpha(f')$ is the noise susceptibility of order $\alpha$. The most general correlator thus reads:
\begin{equation}\label{eq:SaDa}
	\begin{split}
 		\mean{i(f')i(f'')}=\sum_{\alpha\in\mathbb{Z}} &S_\alpha(f') \delta(f''+f'-\alpha f_0)\\
&+ D_\alpha(f')\delta V (f''+f'-\alpha f_0)
	\end{split}
\end{equation}

Because of the adiabaticity,  the quantities $S_\alpha$ and $D_\alpha$ are frequency independent, given by the coefficients of the Fourier series:
\begin{equation}\label{eq:S_fourier}
S(t)=\sum_{\alpha\in\mathbb{Z}} S_\alpha e^{i2\pi\alpha f_0 t},\frac{dS}{dV}(t)=\sum_{\alpha\in\mathbb{Z}} D_\alpha e^{i2\pi\alpha f_0 t}
\end{equation}

\customsubsection{\label{sec:S3env}Environmental thermal noise}

We now evaluate the term $K_{env}$ of \refeq{eq:K3_all}, the contribution to the third moment of the noise generated by the external circuit. Due to the impedance mismatch between the junction and the measuring circuit, the voltage fluctuations propagating from the environment towards the junction are partially reflected back to the amplification chain, and the measured voltage is $v_{mes}(f)=-i(f)\frac{1}{2}t(f)R_J+r(f)v_{env}(f)$ with $r(f)=(Z_{out}(f)-Z_{0})/(Z_{out}(f)+Z_{0})$. The modulation by $v_{env}$ of the variance of the noise generated by the junction leads to a third order correlation:
\begin{equation}\label{eq:Kenv_0}
K_{env}(f)=3 \frac{2 r(f)}{t(f)R_J}\mean{v_{env}(f)i^2(f)}
\end{equation}

\begin{figure}
	\includegraphics[width=0.5\textwidth]{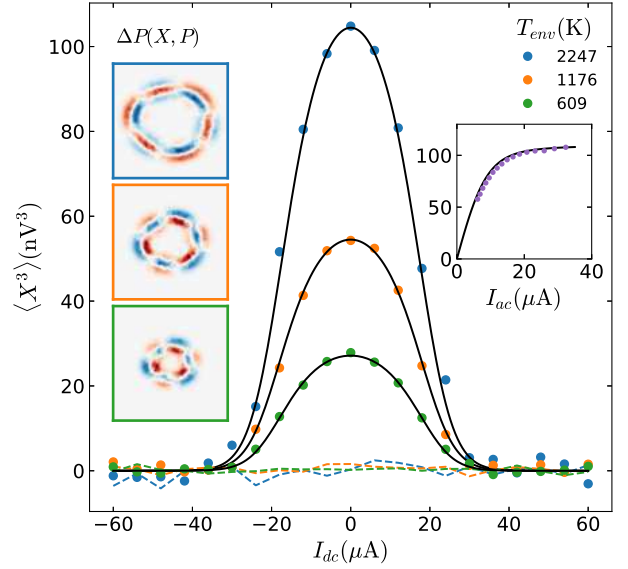}
	\caption{\label{fig:Kenv} $\mean{X^3}$ (solid circles) and $\mean{P^3}$ (dashed lines) for $I_{ac}$=20$\mu A$ in the presence of a tone at frequency $f+83$MHz of various amplitude. Black lines corresponds to fit with \refeq{eq:Kenv}. Quadratures have been rotated to maximize the signal on $\mean{X^3}$. Left insets: differential probability $\Delta \mathcal{P}(X,P)$ for each excitation power. Right inset:  measured $\mean{X^3}(I_{ac})$ for $I_{dc}=0$. }
\end{figure}

\noindent According to \refeq{eq:SaDa}, $\mean{i^2(f)}=D_1(f)\delta V(-f)=D_1(f)t'(-f)v_{env}(-f)$. Since the circuit is reciprocal, $t'(-f)/t(f)=(R_J/Z_0)e^{-2i\phi}$  with $\phi$ the phase of $t(f)$. Introducing the spectral density of the environmental noise $\mean{v_{env}(f)v_{env}(-f)} =\frac{1}{2}k_BT_{env}Z_{0}$  \refeq{eq:Kenv_0} becomes: 
\begin{equation}\label{eq:Kenv}
	K_{env}(f)=3D_1k_BT_{env}r(f)e^{-2i\phi}
\end{equation}

To demonstrate this environmental effect experimentally, we increased the effective temperature $T_{env}$ so that $K_{env}$ becomes the main contribution to the measured third moment. To do so, we excited the sample with a sine wave $v_{env}(t)=A\sin 2\pi(f+\varepsilon) t$ with $\varepsilon = 83\mathrm{MHz}$ in addition to the drive at frequency $3f$. This is equivalent to an increase in the noise temperature at frequency $f+\varepsilon$ in a very narrow band. The effective $T_{env}$ is estimated from the amplitude of the reflected sine wave superimposed to the measured noise. We measured the skewnesses $\mean{X^3}$ and $\mean{P^3}$ for different $I_{ac}$ and $I_{dc}$ and are presented Fig.\ref{fig:Kenv}. One clearly observes a quantitative agreement between the measurements (symbols) and the theoretical prediction of \refeq{eq:Kenv}, both for the dependence on $I_{dc}$ (main plot in \reffig{fig:Kenv}) and $I_{ac}$ (right inset), up to an overall multiplicative factor.

\begin{figure}
	\hspace*{-0.02\textwidth}\includegraphics[width=0.5\textwidth]{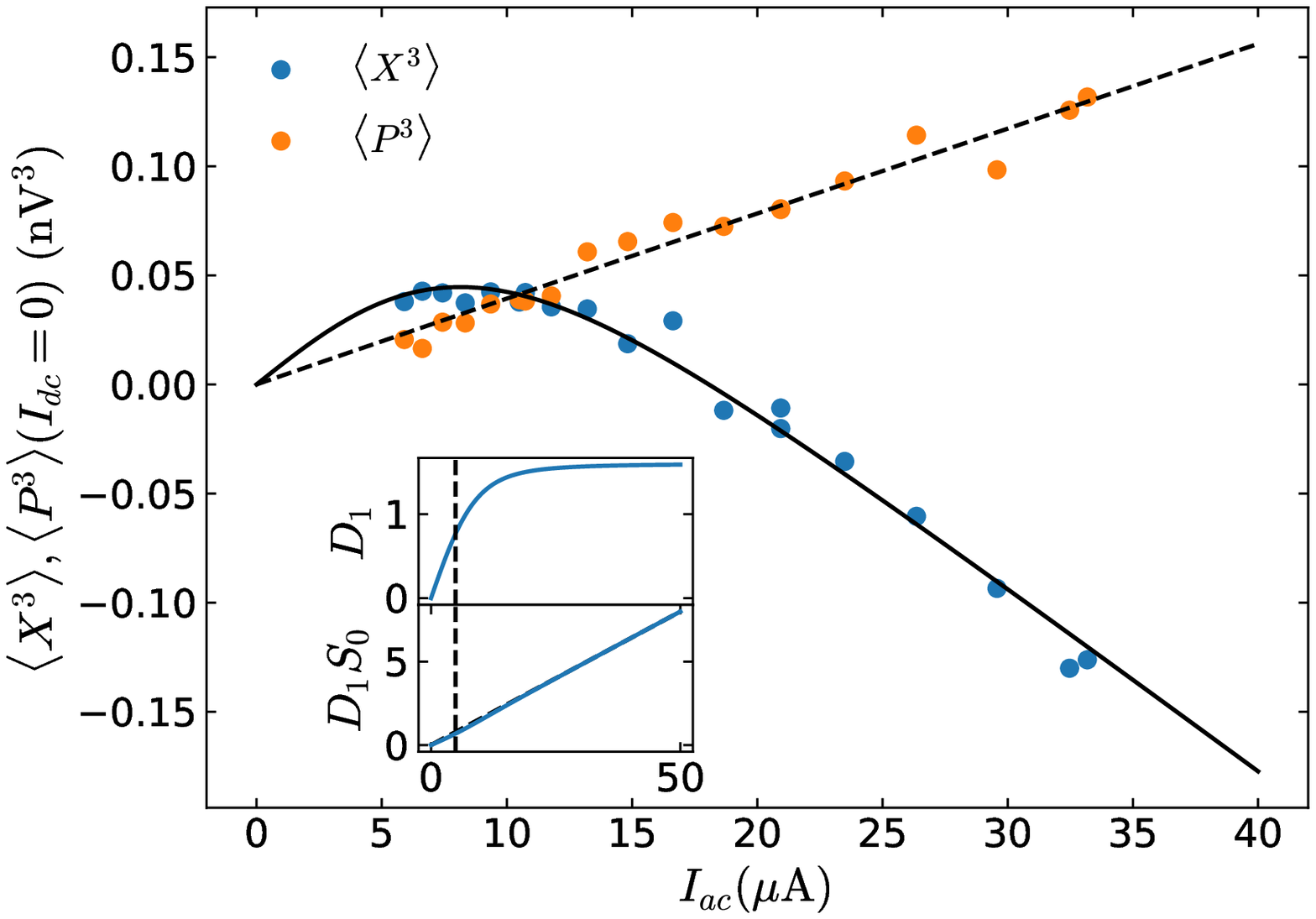}
	\caption{\label{fig:KenvP3} Dependence of the measured third moment with ac bias $I_{ac}$ on both quadratures. Data are fitted by $a D_1(I_{ac})+ bI_{ac}$, with $a=$\powerten{1.2}{11}, $b=$\powerten{-9}{-6} for $\mean{X^3}$ and $a=$0, $b=$\powerten{0.4}{-6} for $\mean{Y^3}$. The global phase has been chosen to maximize contribution of $K_{env}$ on the $X$ quadrature. Insets: Dependence with $I_{ac}$ of $D_1$ and $D_1S_0$ (in arbitrary units). $D_1S_0$ is almost linear whereas $D_1$ shows a saturation for $I_{ac}\gg 2k_BT/eR_J$ (dashed line).}
\end{figure}

\customsubsection{\label{sec:S3fb}Feedback effect}

The feedback term $K_{fb}$ in \refeq{eq:K3_all} represents the contribution of the junction modulating its own noise through the impedance $Z_{in}$. Using \refeq{eq:SaDa} we calculate the feedback contribution to the three current correlation:

\begin{equation}
	\begin{split}
		\mean{i(f)i(f')i(f'')}_{fb}&=\sum_{perm}\mean{i(f)\mean{i(f')i(f'')}}\\
		=\sum_{\substack{perm,\\\alpha\in\mathbb{Z}}} D_\alpha(f')& \langle i(f)i(f'+f''-\alpha f_0)\rangle \Zeff(f'+f''-\alpha f_0)
	\end{split}
\end{equation}

\noindent where the sum is over the cyclic permutation of frequencies $f$, $f'$ and $f''$ with $f+f'+f''=3f$ as imposed by the cyclostationarity. This gives for a narrow-bandwidth:
\begin{equation}\label{eq:Kfb}
K_{fb}(f)=3\sum_{\alpha\in\mathbb{Z}} D_\alpha S_{1-\alpha} \Zeff((2-3\alpha) f)
\end{equation}

\noindent This is a quite unexpected result: the feedback effect on the cyclostationary third moment measured in a narrow band around frequency $f$ involves the environmental impedance not only at $f$ but also at higher frequencies $2f$, $4f$, $5f$, etc. This is a generalization of previous derivation \cite{heikkilaCyclostationaryMeasurementLowfrequency2005}, which treated only the case $I_{dc}=0$.

At zero dc bias $I_{dc}=0$, only the term $D_1S_0\Zeff(f)$ is non-zero. In contrast with the environmental contribution which is proportional to $D_1$ and saturates at high ac excitation, the $D_1S_0$ term is almost linear in $I_{ac}$, see insets of \reffig{fig:KenvP3}. We use this property to separate the different contributions to the skewnesses, which can be reliably fitted by $a I_{ac}+ bD_1(I_{ac})$, with $a$ and $b$ real numbers (which are different for the two quadratures), see \reffig{fig:KenvP3}. Besides the global gain of the measurement, $a=(K_{int}+K_{fb})/I_{ac}$, and  $b=3k_BT_{env}r(f)$. $b$ being known allows us to remove the $K_{env}$ contribution from previous measurements, by subtracting $bD_1(I_{ac}, I_{dc})$ from $\mean{X^3}$ and $\mean{P^3}$ for all values of $I_{dc}$ and $I_{ac}$.

\begin{figure}
	\hspace*{-0.04\textwidth}\includegraphics[width=0.57\textwidth]{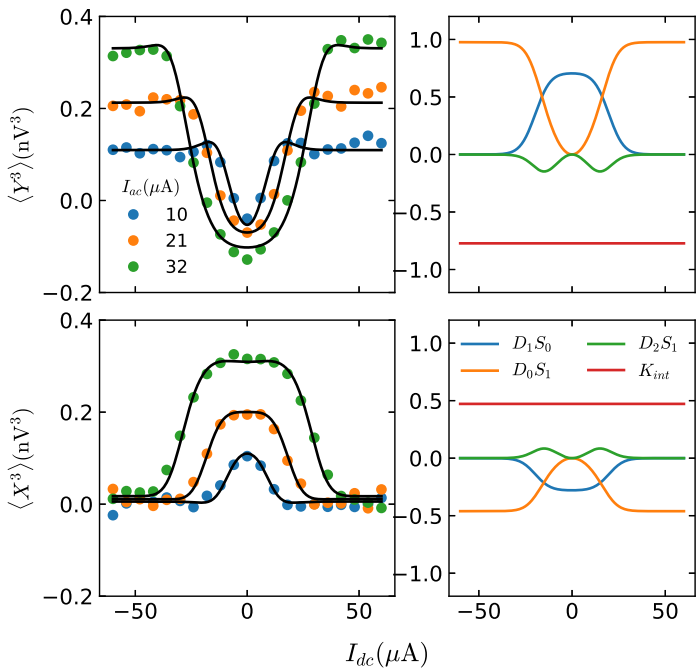}
	\caption{\label{fig:Kfb} Third moment of both quadratures $X$ and $P$ after subtraction of the environmental contribution $K_{env}$. Graphs on the left: Dots represent data and black plain lines are the result from the fitting routine using \refeq{eq:Kfb} plus a constant which is attributed to the intrinsic contribution $K_{int}$. Graphs on the right: each contributions to $\mean{X^3}$ and $\mean{Y^3}$ for $I_{ac}=20\mathrm{\mu A}$ plotted separately.}
\end{figure}

We show in \reffig{fig:Kfb} the skewnesses of both quadratures after subtraction of environmental contributions $K_{env}$. Obviously, these differ from a constant, indicating the presence of extra terms in the feedback contributions, see \refeq{eq:Kfb}. This equation contains an infinite sum involving the environmental impedance taken at very high frequency. We expect however these contributions to decay as frequency increases because of the capacitance of the junction ($C\approx$0.2pF based on geometry) shunting the environmental impedance at high frequency, the cutoff being approximately given by $\sim 20$GHz.

To obtain quantitative results, we extracted the values of the environmental impedance using a fitting routine. First we determine the global gain $G(f)^3\Delta f^2$, from the magnitude of $K_{env}$, assuming $T_{env}=3.7$K and $\abs{r}=0.44$. Then we used a three parameter fit on each quadrature of $K_{int}+K_{fb}$, corresponding to the environmental impedances appearing in $K_{fb}$ for $\alpha < 4$, plus one parameter for the dephasing $\phi$ between $K_{int}$ and $K_{env}$. The fitted data are represented on \reffig{fig:Kfb}. The fitting routine has also be performed with less parameters, by omitting for example the $D_2S_1$ contribution, however the obtained impedances were unrealistic.

%Accordingly, we show in the left graphs of \reffig{fig:Kfb} the theoretical predications when keeping only the terms $\alpha=0$, $\alpha=1$, and $\alpha = 2$ i.e. neglect contributions at frequencies higher than $4f$. While experiment and theory agree well at low ac excitation, there is a clear discrepancy at high $I_{ac}$. The right part of of \reffig{fig:Kfb} shows the difference between the same data after subtraction of the feedback contribution truncated to $\alpha=0,1$ (i.e., solid lines of the left part of the figure). This is very well accounted for by the $\alpha=2$ term, proportional to $Z_{eff}(4f)$, i.e. the environmental impedance at 19.4GHz.

\begin{figure}
	\includegraphics[width=0.5\textwidth]{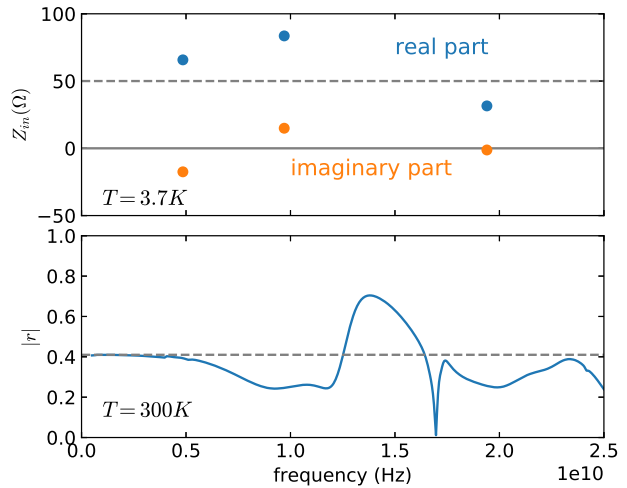}
	\caption{\label{fig:Zin} Top: Input impedance $Z_{in}$ seen by the junction, extracted from the fit of $K_{fb}$ (circles). Bottom: Magnitude of the reflection coefficient of the sample at 300K (solid line), measured using a vector network analyzer. The dashed line corresponds to the reflexion of a pure resistance of 116\ohm, i.e. the dc resistance of junction measured at 300K.}
\end{figure}

\customsubsection{Characterization of the electromagnetic environment}Our data on both quadratures $\mean{X^3}$ and $\mean{P^3}$ are very well accounted for by \refeq{eq:K3_all} for all dc and ac excitations. This involves, beside the total gain $G(f)^3\Delta f^2$, environmental parameters: the electromagnetic response of the environment through the phase $\phi$=-0.1rad (for $K_{env}$) and the impedances $Z_{in}(f)$, $Z_{in}(2f)$ and $Z_{in}(4f)$ (for $K_{fb}$), whose values are presented in \reffig{fig:Zin}.

We also show in \reffig{fig:Zin} the magnitude of the reflection coefficient $|r|$ measured at room temperature with a vector network analyzer. The magnitude below 5GHz is close to what is expected for a transmission line terminated by a simple resistor, which coincides with $Z_{in}(f)$ being close to 50\ohm. At 10GHz, a lower reflection can indicate a better matching which coincides with $Z_{in}(2f)$ being closer to the junction resistance. Our knowledge of the environment is however too crude to predict the value of $Z_{in}(4f)$ from $\abs{r}$.
 
%This involves, beside the total gain $G(f)^3\Delta f^2$, environmental parameters: the environmental temperature $T_{env}=3.7K$ fixed by the temperature of the load of the circulator, and the electromagnetic response of the environment through the phase $\phi$ (for $K_{env}$), and the impedances $Z_{in}(f)$, $Z_{in}(2f)$ and $Z_{in}(4f)$ (for $K_{fb}$). We find for $G^3t^3\Zeff(f)^3\Delta f^2=9.7\times 10^{32}\mathrm{\Omega^3 Hz^2}$, $\phi=$0.04$\mathrm{\pi}$, $Z_{in}(f)=49+i26$\ohm, $Z_{in}(2f)=101+i33$\ohm, and $Z_{in}(4f)=29+i2$\ohm. 

%In order to further confront theory and experiment, we now compare these values with what we expect from the circuit in which the junction is embedded. A simple model is given in the inset of \reffig{fig:Zin}(a): $C$ represents the geometrical junction of the capacitance and $L$ the inductance of the wire bond between the junction contact and the 50\ohm microstrip of the sample holder. We show on \reffig{fig:Zin}(a) the real and imaginary parts of $Z_{in}$ calculated using this model (dashed lines) with $C\approx$0.1pF and $L\approx$1nH (reasonable values given our geometry), as well as the experimental values of $Z_{in}$ given above (full circles). The increase of $Z_{in}$ at $2f=$9.7GHz, is caused by the impedance matching provided by the LC circuit. The model is however too crude to predict the value of $Z_{in}(4f)$.

From this analysis, two conclusions can be drawn for future works. Because of the frequency dependence of reactive part of $\Zeff(f)$ and $r(f)$, all environmental contributions cannot be projected on a single quadrature. Therefore, in order to extract the intrinsic third moment of the sample, both noise quadratures must be measured. Second, a careful design of the electromagnetic environment is necessary to separate the intrinsic contribution from the environmental ones. For example, by providing a low impedance environment at frequency $2f$, $K_{int}$ becomes the dominant contribution at high dc bias. 

\section{\label{sec:Conclusion}Conclusion}
We have measured the cyclostationary third moment of current fluctuations at finite frequency $f=4.85$GHz in a tunnel junction photo-excited at frequency $3f$. We have observed third order correlations between the quadratures of the electric field at frequency $f$ which depends on both the dc and ac bias. Thanks to a theoretical analysis we can decompose these correlations into three contributions: the intrinsic shot noise of the junction, and its modulation by the external noise at frequency $f$ as well as through a feedback mechanism that involves the impedance of the detection circuitry even way outside the detection bandwidth.

Our analysis paves the way towards the design of new experiments probing the third moment of current fluctuations at high frequencies. This includes for example the case of systems with non-trivial dynamics, like diffusive wires \cite{pilgramFrequencydependentThirdCumulant2004} or systems in the regime where the quantum dynamics associated with the timescale $h/eV$ matters \cite{thibaultPauliHeisenbergOscillationsElectron2015,gabelliDynamicsQuantumNoise2008,gabelliNoiseSusceptibilityPhotoexcited2008}. It also includes the case of high impedance samples (quantum dots, Coulomb blockaded systems, etc.) which can be matched to $50\Omega$ only around a single frequency $f$, and where usual wideband techniques fail to operate. It also opens the possibility to use quantum limited amplifiers, such as the Josephson parametric amplifier, to study non-Gaussian noise in the quantum regime. As a matter of fact, the existence of correlations similar to the one we observed but at temperatures such that $k_BT<hf$ would imply third order squeezing in the radiation emitted by a tunnel junction, as has been recently observed in superconducting parametric amplifiers \cite{changObservationThreePhotonSpontaneous2020}. We hope our work will also trigger theoretical advances, such as how to extend theories used to link quantum fluctuations of electrical current to that of the quadratures of the electromagnetic field \cite{grimsmoQuantumOpticsTheory2016}, which has proven successful to predict second order correlations that lead to squeezing, but are probably insufficient to account for third order correlations.

We acknowledge fruitful discussions with Edouard Pinsolle, and the technical help of G. Lalibert\'e. This work was supported by the Canada Excellence Research Chair  program,  the  NSERC,  the  MDEIE,  the  FRQNT via  the  INTRIQ,  the  Universit\'e  de  Sherbrooke via the EPIQ, the CFREF via  the Institut Quantique and the Canada Foundation for Innovation.

\bibliography{references}
\end{document}